# Extensive fitness and human cooperation

J. H. van Hateren

Johann Bernouilli Institute for Mathematics and Computer Science,
University of Groningen, The Netherlands

e-mail: j.h.van.hateren@rug.nl

**Abstract** Evolution depends on the fitness of organisms, the expected rate of reproducing. Directly getting offspring is the most basic form of fitness, but fitness can also be increased indirectly by helping genetically related individuals (such as kin) to increase their fitness. The combined effect is known as inclusive fitness. Here it is argued that a further elaboration of fitness has evolved, particularly in humans. It is called extensive fitness and it incorporates producing organisms that are merely similar in phenotype. The evolvability of this mechanism is illustrated by computations on a simple model combining heredity and behaviour. Phenotypes are driven into the direction of high fitness through a mechanism that involves an internal estimate of fitness, implicitly made within the organism itself. This mechanism has recently been conjectured to be responsible for producing agency and goals. In the model, inclusive and extensive fitness are both implemented by letting fitness increase nonlinearly with the size of subpopulations of similar heredity (for the indirect part of inclusive fitness) and of similar phenotype (for the phenotypic part of extensive fitness). Populations implementing extensive fitness outcompete populations implementing mere inclusive fitness. This occurs because groups with similar phenotype tend to be larger than groups with similar heredity, and fitness increases more when groups are larger. Extensive fitness has two components, a direct component where individuals compete in inducing others to become like them and an indirect component where individuals cooperate and help others who are already similar to them.

**Keywords** Inclusive fitness · Extensive fitness · Agency · Cooperation · Human evolution

## Introduction

One of the defining characteristics of the human species is its capacity to establish cooperation within large groups of unrelated individuals. From an evolutionary point of view, this may seem somewhat puzzling. Whereas cooperation between genetically related individuals can be readily understood from inclusive fitness theory (in particular kin selection), additional mechanisms are required when individuals are unrelated. In recent decades, several such mechanisms have been proposed (reviewed in Rand and Nowak 2013), such as direct reciprocity (Trivers 1971), indirect (reputation-based) reciprocity (Nowak and Sigmund 2005), and multilevel selection (Wilson 1975). The combined force of such mechanisms is thought to explain not only the extent of human cooperation, but also the path by which it has evolved and the ways by which it is stabilized against cheating.

In this article, I offer a fresh view on the evolutionary value of human prosociality. This view is a further elaboration of a recently developed theory on the origin of agency and meaning (van Hateren 2014b, 2015a). The theory assumes that organisms have evolved an internalized, estimated version of their actual external fitness (their expected rate of reproducing). The internalized version, implicitly present as a distributed process throughout the organism's physiology, is the one that produces goal-directedness and meaning intrinsic to the organism. It is also responsible for those parts of the organism's behaviour that involve agency. I propose here that in humans, the external and internalized versions of fitness have evolved into a variant that extends beyond inclusive fitness – a variant that is called 'extensive fitness' here. Whereas inclusive fitness can enhance fitness through indirect effects affecting individuals that are genetically similar (such as kin), extensive fitness can also enhance fitness through indirect effects affecting individuals with similar phenotype. Computations on a simple model that is presented below suggest that such a mechanism is evolvable. Extensive fitness comes in two forms, a direct and indirect one, of which the former implies competition between individuals and the latter cooperation. It is argued that the cognitive requirements of the mechanism are such that it is probably only present in its full-blown form in humans, although it might be present in primordial form in other species as well.

The theory presented here focusses on the effects of internalized fitness, arguing for its unique role in human cooperation. Because internalized fitness is a novel concept, the theory differs significantly from theories focussing only on conventional fitness (e.g., Frank 2003; Queller 2011) or on fitness based on cultural transmission (e.g., El Mouden et al. 2014). However, there is no conflict here, because the various mechanisms can coexist and reinforce each other. Nevertheless, when agency is involved in cooperation, it is conjectured that internalized fitness must be involved as well (van Hateren 2014b).

The article is organized as follows. "A simplified model with heredity and behaviour" explains the basic model that has been used before for the simplest form of fitness, direct fitness. Although the section contains no new material, it is needed for understanding the subsequent extensions to inclusive and extensive fitness. Readers familiar with this material may wish to go directly to "Incorporating inclusive fitness". That section explains how inclusive fitness can be incorporated into the model by introducing a nonlinear factor that enhances fitness depending on the size of a subgroup of individuals with similar hereditary properties. In "Incorporating extensive fitness", it is shown that applying a similar fitness enhancement to subgroups of individuals with similar phenotypic behaviour produces populations that outcompete populations only incorporating inclusive fitness. I discuss the consequences of the theory for competition and cooperation in the "Discussion". Finally, the "Conclusion" recapitulates the main findings of the article. Mathematical details of the model are in the "Appendix".

## A simplified model with heredity and behaviour

In this section, I will summarize the theory and model that have been more extensively explained in van Hateren (2015a). The description will be qualitative here, with a mathematical description and details in the "Appendix". An essential component of the model is a special form of causation, called modulated stochastic causation (Fig. 1a). In modulated stochastic causation, a non-negative deterministic variable (upper trace in Fig. 1a, taken as a function of time) drives the variance of a stochastic (random) process (lower trace). The deterministic variable (which may represent a system state or property) is itself caused by other factors (arrow 1), and the resulting stochastic variable subsequently drives other variables (arrow 3). The causation here is neither completely deterministic (because of the stochasticity), nor completely stochastic (because the lower trace is not completely random, but has its variance varied in a deterministic way).

Modulated stochastic causation is assumed to play a role on at least two different timescales relevant for living organisms. First, the timescale of hereditary change, which occurs on an evolutionary timescale along a specific line of descending organisms. Second, the timescale of behavioural change, which involves changes of the organism's phenotype (i.e., its actual form as confronting the world) during an organism's lifetime. For simplicity, behaviour is defined here in the broadest possible sense, including any physiological changes within an organism. It includes development and learning, and it may be equated with phenotypic plasticity if that is defined broadly (e.g., Snell-Rood 2013). The right half of Fig. 1b sketches how modulated stochastic causation affects hereditary and behavioural change. But before that is explained, we will first focus on the left half of the diagram, which represents the basic theory of Darwinian evolution.

The organism in Fig. 1b is assumed to be embedded in a time-varying environment, including other organisms. This environment varies continually and (partly) unpredictably over a wide range of timescales (Bell 2010; van Hateren 2015a), spanning those of evolution as well as those of behaviour. The organism has a fitness, called $f_{true}$ below, resulting from the organism's characteristics and its interactions with the environment. Fitness $f_{true}$ is defined here as a predictive variable that gives the (expected) rate of reproduction, i.e., the expected number of offspring per unit of time. For each organism, fitness is thus a continuous function of time, quantifying how well the organism is doing – as an expectation, probabilistically, with what is actually realized deviating stochastically from the expectation. When circumstances deteriorate, e.g. during a famine, fitness $f_{true}$ decreases, down to zero if the organism dies. But it can increase again if the organism survives and circumstances improve. It should be stressed that fitness as used here is a predictive variable, not equivalent to the actual number of offspring an organism gets. The latter is just the stochastic realization of fitness as accumulated over an organism's lifetime. Whereas $f_{\text{true}}$ itself is taken to be a determinate variable, with a well-defined value varying in time, it produces a stochastic outcome for individuals (as actually realized offspring). The outcome is stochastic because of the complexity of individual and environment, and because they are presumably sensitive to inherently random factors (such as thermal noise, quantum noise, and external influences from unknown and untraceable sources).

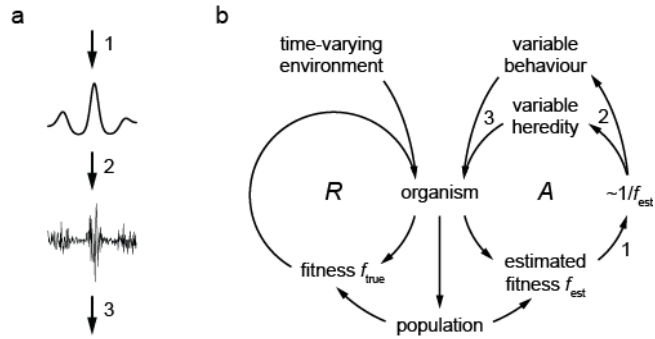

**Fig. 1** The role of $f_{\text{est}}$ in generating hereditary and behavioural variability. **a** In modulated stochastic causation, a time-varying non-negative deterministic variable (upper trace) is caused by other variables (arrow 1) and drives (arrow 2) the variance of a stochastic (random) process (lower trace), which then drives other variables (arrow 3). **b** Extended mechanism of Darwinian evolution. In the reproductive $R$ loop an organism (probabilistically) reproduces in proportion to its fitness, $f_{\text{true}}$, which depends on a time-varying environment and on the properties of organism and population. Hereditary and behavioural variability are varied as driven by an internal estimator of fitness, $f_{\text{est}}$, implicitly made in the organism itself. Variability is high when $f_{\text{est}}$ is low, and low when $f_{\text{est}}$ is high (symbolized by $\sim 1/f_{\text{est}}$). The active $A$ loop then continually utilizes modulated stochastic causation (arrows 1-3 as in **a**), and produces a genuine goal-directedness and an elementary form of agency.

It is important to emphasize that $f_{\text{true}}$ as defined here is not a simple parameter, but rather a complex process with many inputs from environment and organism. It has an intricate internal dynamics (that is, a structure or form) and one output (the rate of reproduction). This way of defining $f_{\text{true}}$ is somewhat analogous to a typical mathematical function, which also has a value, a form, and argument(s). The difference is that the way fitness is produced in nature is dynamic and also depends on memory – genetic and behavioural memory as stored in organisms, as well as environmental memory as produced, for example, by niche construction (Odling-Smee et al. 2003; Laland and Sterelny 2006). When fitness is discussed below (either $f_{\text{true}}$ or its internal estimate $f_{\text{est}}$) then this refers either to a process or to its outcome, a value giving a rate of reproduction. Usually this will be clear from the context, but where it is ambiguous, I will specifically refer to it as the value or form of fitness.

A second point to emphasize is that fitness is defined here as an individual property. But this does not exclude population effects, because such effects can always be incorporated into the fitness process as a population-level feedback onto individuals. For example, the size of the population will control individual fitness when resources become scarce (the Malthusian factor). Also frequency-dependent fitness is readily modelled as a nonlinear feedback. Genetic and physiological effects on individual fitness are automatically incorporated, because they are just part of the form of the fitness process. It should be noted that fitness as used here models the way it is produced in nature, but it is not directly observable other than through its stochastic outcome. Constructing and testing an accurate model would require considerable effort even in the case of simple organisms in a controlled environment. Simplified, approximate models would be more feasible. Nevertheless, $f_{\text{true}}$ as defined here is an indispensable and powerful theoretical tool for understanding the extensions of evolutionary theory discussed below.

For a large subpopulation of similar organisms, fitness as defined above can be used, with suitable averaging, to predict how the subpopulation will grow. When fitness is, on average, above the replacement level, the subpopulation is expected to grow explosively (exponentially), until growth is limited by environmental constraints. On the other hand, fitness below the replacement level increases the likelihood of eventual extinction. As mentioned above, the model of Fig. 1b incorporates population effects as feedback onto the fitness of individual organisms. Such effects may be seen as arising from the environment, if that is defined broadly as including all other organisms. However, for the sake of clarity, Fig. 1b shows 'population' explicitly. The arrow from organism to population stands for the fact that organisms produce the population and the arrow from population to fitness symbolizes how the population affects the individual. Population effects other than the standard Malthusian factor will be added below when inclusive and extensive fitness are incorporated into the model.

The basic mechanism of Darwinian evolution by natural selection is represented by the reproductive loop $R$ in Fig. 1b. Organisms get offspring that inherit similar but not identical properties – descent with (stochastic) modification. Such properties then combine with environment and population size to produce a fitness in the offspring. Offspring with typically high fitness is likely to reproduce more than offspring with lower fitness. In this way, the properties of the individuals in the population may gradually change, depending on the time-varying environment and the stochastic hereditary changes. Environmental factors that remain sufficiently stable over many generations can thus lead to organisms with well-adapted properties (natural selection). For simplicity, the description of evolutionary change is kept as basic as possible here. However, the theory presented below would work equally well when mechanisms that are more complex (see, e.g., Pigliucci and Müller 2010) would be incorporated (see also van Hateren 2014a, for more discussion).

In the basic theory of Darwinian evolution, the hereditary variation that occurs in offspring arises from a stochastic process with a variance that is fixed or at least not the target of continuous modulation by the organism itself. There are various sources of such variance in nature, such as genetic mutation, recombination, and sex, all summarized here by the term 'hereditary variability'. As discussed elsewhere (van Hateren 2013, 2014a, 2015a,b), the basic theory is extended by letting the hereditary variability depend on an internal estimate made within the organism itself of its own fitness, called $f_{est}$ below. 'Estimate' is used here in the technical sense (as in estimation theory), and does not indicate anything deliberate or intentional. Moreover, the $f_{est}$ used here is intrinsic to the organism and does not refer to a human-made empirical estimate of fitness, such as by counting offspring or by more sophisticated methods (e.g., Jost 2003). Where $f_{true}$ is a theoretical, expected reproductive rate, $f_{est}$ is an estimate of that expected rate. Both processes produce a fitness (and estimated fitness) before the actual outcome is realized.

The internal $f_{est}$ refers to an evolved process that is implicitly present in the organism's physiology, presumably in a distributed form and depending on a large set of fitness indicators that are available to the organism. For example, specialized sensors may detect external threats or opportunities in the form of adverse substances or nutrients. The internal state of the organism can be similarly monitored. More advanced indicators are possible in organisms with advanced nervous systems. Note that $f_{est}$, like $f_{true}$, has both a value and a form. The form is in this case an intrinsic physiological process present in the organism, and the output of this process, its value, is the estimated rate of reproduction. This output is most likely not an explicit one, such as in the form of a single physiological variable, but rather an implicit one, distributed throughout the process.

The extended mechanism works as follows (right half of Fig. 1b). When fitness is above the replacement level, the hereditary and behavioural variability should be low because things are going well and they are expected to keep going well if the environment does not change too much. Because environments are assumed to contain strong components that indeed change only slowly, this is a reasonable expectation. On the other hand, when fitness is below the replacement level, not changing the organism's properties would be a poor strategy with a high risk of eventual extinction. Variability should therefore be increased. Although this will lead to many variants that have even lower fitness (and thus are likely to become extinct even faster), the occasional variant with higher fitness can overrule this disadvantage. This is because variants with high fitness can grow exponentially in numbers. The potential for exponential growth can compensate, on average, for the low probability of favourable variants. Simulations on simple models like the one discussed in the "Appendix" show that populations with modulated variability outcompete populations with fixed variability (van Hateren 2015a). This indicates that the mechanism of modulating variability is evolvable, at least in principle. In addition to hereditary variability as relevant for evolutionary timescales (van Hateren 2014a), behavioural variability can perform a similar – and evolvable – role at behavioural timescales (van Hateren 2015a). For behaviour, the drive is not so much instant, actual exponential growth, but the expectation of exponential growth (recall that fitness is the expected rate of reproduction, varying from moment to moment). The mechanism is beneficial on average, statistically.

In particular for behaviour, variants will usually not be completely random. They will vary randomly on top of known behavioural change that is expected to work reasonably well as established by prior natural selection or learning. Such known beneficial strategies should obviously be performed by the organism, automatically. No stochastic modulation is needed for that. Thus the *A* loop of Fig. 1 only concerns those changes of which the consequences cannot be foreseen, which are in fact those parts of behaviour that involve agency rather than automaticity (van Hateren 2014b). For simplicity, this separation of behaviour into parts with foreseeable and unforeseeable effects is not further elaborated upon here. It could in principle be included in the computational model presented below as a crude time-varying offset of the environmental variable, foreseeing part of this variable and, in effect, reducing the environmental variance. Alternatively, the environmental variable as actually used in the computations could be interpreted as the residual after such an offset has already been accounted for.

In Fig. 1b, $\sim 1/f_{est}$ symbolizes the dependence of variability on fitness (small variability when fitness is high and large variability when fitness is low). However, the actual form is under selection pressure and can be somewhat different. Furthermore, the theory assumes that the internal fitness estimate made by the organism is indeed close enough to $f_{true}$ for the mechanism to work. This internal estimate is in fact also under selection pressure, that is, it is driven to be as close to $f_{true}$ as possible, given the means available to the organism. The closer $f_{est}$ is to $f_{true}$, the better the mechanism works. However, in real organisms, this would need to be balanced against the fitness costs of putting more effort into making a good estimate.

Although $f_{est}$ focusses its many inputs onto a single output (the estimated rate of reproduction), this single output must subsequently be expanded again into a range of possible hereditary and behavioural changes. The simple model used for the calculations presented here requires no such expansion, because it is deliberately formulated as one-dimensional – heredity, behaviour, and environment all lie along a single dimension. However, in the more realistic multi-dimensional case, not all possible dimensions should be varied equally much. Particularly for behaviour, the partial fitness effects of the various inputs and behavioural outputs, and how they are correlated, should be taken into account and should be properly weighted. This transformation of $f_{est}$ into various behavioural changes is, like $f_{est}$ itself, expected to be a process distributed throughout the organism's physiology. Constructing it from first principles would be extraordinarily difficult in realistic cases, but there is no reason why it could not readily evolve through standard evolutionary mechanisms.

The arrows 1-3 in Fig. 1b correspond to the arrows 1-3 in Fig. 1a. In other words, the *A* loop of Fig. 1b incorporates modulated stochastic causation. The deterministic factor corresponds to $f_{est}$, which is itself a determinate, well-defined rate, like $f_{true}$. The stochasticity modulated by $f_{est}$ corresponds to hereditary and behavioural variability. These variabilities have been drawn together in Fig. 1b for simplicity (arrow pairs 2 and 3), but the two branches actually have different timescales (see the "Appendix"). Because the *A* loop is a feedback loop (i.e., a loop displaying cyclical causation), the deterministic and stochastic components become inseparably entangled. The value of $f_{est}$ drives the variability, the variability determines the probability of specific hereditary and behavioural outcomes, the resulting heredity and behaviour interact with the environment and produce a new $f_{true}$ and therefore a new $f_{est}$, the new $f_{est}$ drives again the variability, and so on and so forth. The result of this process is a genuine form of goal-directedness and an elementary form of agency, as will be explained now.

The *A* loop will result, on average, in high $f_{est}$, purely for statistical reasons. The hereditary and behavioural variation will effectively search through hereditary and behavioural space – and will even construct new parts of that space. Dispersion away from areas in that space with low fitness will be quick because of the high variability there. In contrast, dispersion away from areas with high fitness will be slow because of the low variability there. As a result, areas with high fitness will get a higher probability of being occupied, with many individuals of a population present (for heredity) or much time spent there (for behaviour). High $f_{est}$ thus effectively functions as a goal for the organism, independently of what it represents (for a computational example of attaining an arbitrary goal see figure 1b in van Hateren 2015a). However, the only goal that would be stable on an evolutionary timescale is fitness itself, and therefore $f_{est}$ must represent $f_{true}$. High $f_{true}$ in isolation is not a real goal, because evolution has no foresight: high $f_{true}$ is merely the consequence of a conventional process that occurs naturally whenever there is a resource-limited capability to reproduce with small changes. In contrast, by internalizing $f_{true}$ as $f_{est}$, the organism obtains high $f_{est}$ as a genuine goal (van Hateren 2015a). It is an evolved property, produced by the special stochastic feedback loop sketched in Fig. 1b, in combination with the evolutionary pressure on $f_{est}$ to align itself with $f_{true}$. It should be noted that high $f_{est}$ is the ultimate overall goal of an organism, but in practice, this must be subdivided into a range of more specific sub-goals. On average, such sub-goals should contribute to the ultimate goal. How sub-goals can be generated and can be kept aligned with the ultimate goal is clearly a highly complex topic that is beyond the scope of this article.

The *A* loop leads to hereditary trajectories (consecutive hereditary properties of organisms along a line of descent) and to behavioural trajectories (consecutive behavioural properties of an organism). Such trajectories combine the spontaneity of the stochasticity in the loop with the deterministic goal-directedness of $f_{est}$. Because the spontaneity and goal-directedness are entangled and cannot be separated, this establishes for behaviour a primordial form of agency: some behavioural freedom in combination with a certain deliberateness (van Hateren 2014b).

Individual fitness forms the core of the theory summarized here. Agency can only be understood as a property originating from individual organisms, with $f_{est}$ as the key component. In previous work (van Hateren 2014b, 2015a), fitness was mostly described as direct fitness, i.e., involving direct reproduction. Variants of fitness that incorporate more complexity, such as caused by genetic relatedness and through behavioural mechanisms, were mentioned but not analysed computationally. Below such an analysis will be performed by elaborating on the model in Fig. 1b (as detailed in the "Appendix", section "Inclusive and extensive fitness").

## Incorporating inclusive fitness

The model analysed in van Hateren (2015a) quantifies the heredity of an organism by a single parameter *h* and its behaviour by a single parameter *b*. The result of an organism with heredity *h* and behaviour *b* – at a particular point in time – is a phenotype (i.e., a phenotypic behaviour) $p=h+b$. Heredity and behaviour are here taken to have the same units. Although this model is highly simplified compared with real organisms, it has the advantage of being computable and understandable. Moreover, it is still sufficiently complex for producing interesting insights as will become clear below.

The organism is part of a population of organisms distributed over a range of values of *h* and *b*. It is embedded in a time-varying environment *E* (also a single parameter, with the same units as *h* and *b*). An organism's fitness $f_{true}$ depends on the values of *E*, *h*, *b*, and the total number of organisms in the population. This total number may vary in time, depending on how well the population is adapted to *E*. The fitness is defined to be smaller when *E* and *p* differ more and also when the size of the population is larger. Ideally, when *E* changes, *p* should follow such a change and become identical in value to *E* (by changing either its heredity *h* or its behaviour *b*, or both). As a simple example, *h* might stand for the inherited weight of organisms, *b* for their behavioural weight adjustment (e.g., by eating less or more), *p* for their resulting actual weight ($h+b$), and *E* for the (time-varying) environmentally preferred weight. The difference between *E* and *p* should ideally be zero for maximum fitness.

Behaviour *b* can vary much faster than heredity *h*, but *h* is assumed to cover a larger range of possible environments (values of *E*) than *b* could cover at any point in time. Adaptation to new *E* therefore involves fast (behavioural) adaptation – by modifying *b* – in combination with slow (hereditary) adaptation – by modifying *h*. Figure 2a shows a snapshot of a simulation of how a population of organisms continually adapts to varying *E*. The black dot gives the value of *E* at the particular point in time shown (with *E* at about 52). The two curves show how the population is distributed over individuals with different values of *h* (black line) and with different values of *h+b* (grey line). The two curves each contain the same population, but as can be seen, the *h+b* curve matches the current value of *E* more closely. This merely shows that *b* performs adaptation in addition to *h*. In van Hateren (2015a), computations are presented that show that populations having both *h* and *b* outcompete populations having only *h* (i.e., when there is no behavioural variation at all). The widths of the *h* and *h+b* curves depend on the base variability rates in the model (on top of which

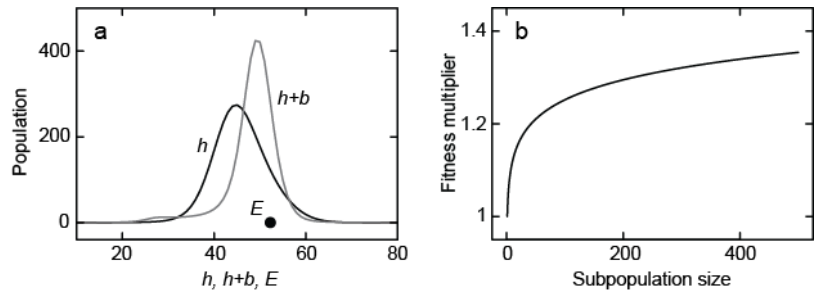

**Fig. 2 a** Snapshot of the simulated evolution of a population with varying heredity $h$, varying behaviour $b$, and only direct fitness. The environment $E$ varies continually and unpredictably. At the time shown, the value of $E$ is given by the position of the black dot along the horizontal axis, the distribution of heredity in the population is given by the black curve marked '$h$', and the distribution of phenotypes ($p=h+b$) is given by the grey curve marked '$h+b$'. In the simulation, $b$ can vary 100 times faster than $h$ (see the "Appendix"), which results in $h+b$ typically being closer to $E$ than $h$. The distributions are drawn here as continuous curves for the sake of clarity, but $h$ and $b$ (but not $E$) are restricted to integer values. **b** Inclusive and extensive fitness are implemented by giving subpopulations of size $n$ a fitness advantage by multiplying fitness by $g^{\log(n)}$ (shown here as a function of $n$ for $g=1.05$). Subpopulations are defined by heredity $h$ for inclusive fitness, and by phenotype $p=h+b$ for the phenotypic part of extensive fitness (which incorporates inclusive fitness as well).

heredity and behaviour are varied through $\sim 1/f_{est}$). The curves could be narrower (and perhaps closer to a particular value of $E$) when these rates would be lower. However, when rates are too low the population would not be able to follow the occasional fast changes in $E$ and would quickly become extinct. The curves could be broader when the base rates would be higher, but then the mean distance to $E$ would be larger. When rates get too high, the resulting lower mean fitness could then also lead to extinction, or at least to a disadvantage relative to other populations that are better adapted to the statistics of $E$.

Fitness in its simplest form (direct fitness) only involves direct offspring, such as would be relevant for an asexually reproducing species without any form of gene exchange between same-generation individuals. However, sexual reproduction and horizontal gene exchange are very common in nature, and direct fitness then needs to be replaced by inclusive fitness (Hamilton 1964). Inclusive fitness incorporates the indirect fitness gain an individual can obtain by helping to increase the fitness of genetically related individuals. The basic idea is that from the point of view of the genes of an individual, it is irrelevant how the probability of their presence in next generations is increased. The probability could be increased by direct offspring of the individual itself, or indirectly, through individuals with related genomes. The probability of transfer to the next generation may vary, depending on relatedness, but as long as helping related individuals increases the overall probability, it may be worthwhile (depending on the costs). Note that I am using the term 'inclusive fitness' here in the general sense of broadening direct fitness through genetic relatedness. Its use here is therefore agnostic to the recent debates on the usefulness of inclusive fitness in its more precisely defined forms (e.g., Nowak et al. 2010; West and Gardner 2013). The current approach also does not require a specification of the level of selection, because fitness is equated here with the entire process leading to a rate of reproduction. Therefore, it automatically includes all relevant levels and it might be seen as, inherently, a multi-level approach.

For the model as used thus far, the mode of reproduction is implicitly asexual and there is no horizontal exchange of heredity. Nevertheless, the individuals in the population that have a particular value of $h$ have identical hereditary properties, by definition. Individuals with other $h$ have different hereditary properties, again by definition. From the point of view of the heredity of an individual, it is irrelevant whether such heredity is transferred directly to offspring, or indirectly through other individuals with the same $h$. Helping such other individuals to increase their direct fitness can therefore be a valid strategy if it increases the overall productivity of type $h$. This argument is similar to the one given above to explain the indirect parts of inclusive fitness. In a way, all individuals of type *h* are 'relatives' of each other. If that is so, then how should one adjust their fitness in order to incorporate indirect fitness?

One possibility is to use an additive model, i.e., a small amount of fitness is added to each pair of interacting (helping) relatives. The added fitness can then be understood as the net result of both subtracting and adding some fitness (Hamilton 1964; Grafen 2006; West et al. 2011). Some fitness should be subtracted, because the helper presumably decreases its own direct fitness by helping relatives (because helping takes time and energy). Some fitness should be added, because the increased fitness (of the individual being helped) that results from the helping should be added to the helper's fitness (discounted by the relatedness of the relative, i.e., as in Hamilton's rule, see e.g. Gardner et al. 2011; but that relatedness is constant here). Another, equivalent way of accounting just subtracts the fitness cost of helping and adds the fitness benefit of being helped, assuming symmetrical helping on average.

However, for the current models, subtracting and adding fitness in this way can lead to invalid results. For example, a negative fitness could result if an individual would help many others in an ineffective way (always subtracting a little of its own fitness, but never adding fitness). But fitness is a reproductive rate (expected number of offspring per unit of time), and therefore non-negative. For a non-negative quantity, a multiplicative model is more appropriate. In such a model, the fitness of both partners is multiplied by a factor slightly larger than unity for each pair of interacting relatives (assuming reciprocated help). The more relatives there are, the larger the number of possible interactions, and the larger the overall multiplication factor. However, if the group of relatives becomes large, it is not realistic to assume that all possible interactions are fully utilized. The fitness benefits will grow slower and slower with the number of relatives. This is incorporated in the model by the way fitness is increased as a compressive (sublinear) function of the number of relatives, that is, of the size of the subpopulation of type *h* (Fig. 2b).

The fitness multiplier has, qualitatively, the same role as the balance between the benefits and costs of helping in the model of Hamilton (1964). The larger the multiplier, the more the organisms benefit from the cooperation. The fitness multiplier can be understood to arise from symmetrical, reciprocated helping for each pair, but alternatively it can be given a statistical interpretation. For the latter it does not assume symmetry for each interaction, merely symmetry in the sense of balancing the probability of helping an arbitrary member of the *h*-subpopulation with the probability of being helped by an arbitrary (possibly other) member of the *h*-subpopulation.

This way of incorporating inclusive fitness is another population feedback onto individual fitness, like the Malthusian factor mentioned above. But now it increases rather than decreases fitness, and it does not depend on total population size, but only on the size of the specific subpopulation for each type *h*. It is analogous to the frequency-dependent effects on the fitness of specific traits that are well known in evolutionary biology (e.g., for various forms of mimicry, where the effectiveness of a mimicry depends on how many individuals have that trait).

Simulations where a population incorporating inclusive fitness in this way is made to compete with a population without it show that the former always outcompetes the latter. Competition is simulated by letting the two populations start with equal size and share the same environment (with carrying capacity $K$, see Eq. 4 in the "Appendix"). Fitness of all organisms in each population is then reduced according to the Malthusian factor $\exp(-N(t)/K)$, with $N$ the total number of organisms in both populations. By using environmental resources, each organism's presence negatively affects the fitness of all others by contributing to $N$ and thereby increasing the fitness reduction caused by the Malthusian factor. This results in general competition, giving an advantage to types of organisms that have higher fitness, on average. For 100 different simulations, Figure 3 shows the ratio of the number of organisms in populations without and with inclusive fitness (black line: mean, grey lines: mean ± standard deviation). Organisms with inclusive fitness outcompete organisms with only direct fitness. This result is expected, because inclusive fitness (*h*-helping) is specified here as always increasing direct fitness (through Fig. 2b).

A complication that is not addressed here, but that will be important in practice, is that organisms need a way to identify which other organisms are of the same hereditary type (or at least a way to have similarity expected, such as through proximity, Hamilton 1964). Errors in identification, the presence of impostors (non-relatives pretending to be relatives), and freeloading (consistently accepting more help than giving), will all reduce the effectiveness of the mechanism. Nevertheless, when such problems are controlled to some extent, the mechanism may still remain positive for fitness, on balance.

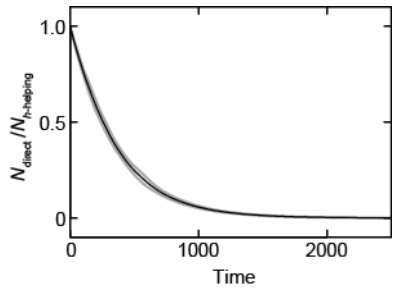

**Fig. 3** Populations with *h*-helping outcompete populations with only direct fitness. The lines show the ratio of the number of organisms in a population with direct fitness ($N_{\text{direct}}$) and one with *h*-helping ($N_{h\text{-helping}}$), as a function of time (100 time steps correspond to an organism's lifetime). Black line: mean of 100 simulations with different realizations of the environment $E(t)$, grey lines: mean ± standard deviation.

**Incorporating extensive fitness**

Whereas the fitness gain for inclusive fitness depends on the size of the subpopulation with the same heredity (*h*), another possibility is to let an organism's fitness depend on the size of the subpopulation with the same phenotype (*p*). As defined above, the phenotype is the result of heredity *h* as modified by behaviour *b*, as $p=h+b$. It is the observable behavioural phenotype of organisms. Identifying which organisms belong to a specific subpopulation *p* may be feasible for organisms with large brains. The idea is that the members of subpopulation *p* help each other in the same way as above for members of subpopulation *h*. If the type of helping is the same, the fitness gain should follow the same curve as before (Fig. 2b). A possible complication here is that there is a considerable chance that the individual being helped (B) is of another type *h* than the helper (A). The fitness gain of B then does not contribute to A's hereditary fitness, not even partly. However, there is an equal chance that A is being helped itself by an unrelated individual (C), and A's fitness gain should then similarly not contribute to C's hereditary fitness, not even partly. Instead, it should be attributed to A's fitness. The net effect of helping is therefore the same as for *h*-based subpopulations, and it should indeed follow the same curve. Another, more abstract way to see this equivalence goes as follows. The total gain in fitness of the entire population only depends on the total amount of helping behaviour, not on the specific *h* or *p* of the helping pairs. This fitness gain must then become distributed in a way that only depends on how often an individual is part of a helping pair (for lack of any other criterion in the model), and thus produce the same curves for *h*- and *p*-helping. As before, interactions could involve either direct reciprocation, or only reciprocation in a statistical sense.

What would happen if two populations, one based on helping relatives (*h*-subpopulations) and the other based on helping similar phenotypes (*p*-subpopulations) were made to compete with one another? At first sight, one might expect trouble for the *p*-helping. The reason is that evolution depends on heredity *h* for long-term adaptation, and *p*-helping is 'misfiring' in the sense of often helping individuals with a type *h* different from the helper's type. In contrast, *h*-helping is not misfiring, but it helps those with the same hereditary properties. One might therefore expect that *h*-helping will eventually outcompete *p*-helping. Figure 4a shows the typical result of such a simulated evolution. As before, two populations of equal initial size are competing, with one population using *h*-helping (grey trace) and the other population using *p*-helping (black trace). Surprisingly, *p*-helping outcompetes *h*-helping, within about 40 generations (the typical lifetime of the organisms in this simulation was 100 time steps). This was found consistently for each of 100 simulations performed with different realizations of the time-varying environment (Fig. 4b).

What is going on? The first point that needs to be understood is that the 'misfiring' argument is wrong. For natural selection, *p* is as important as *h*. Whereas indeed *h* is the one with memory, *p* is the one that confronts the environment. Both are needed. It is true that *p*-helping misfires in some sense, because the *h* that is actually helped is likely to be different from the helper's *h*. This might transfer heredity that is, on average, poorly adapted. But at least a *p* is helped that is matched to that of the helper, and if there are many helping pairs of type *p* then such a *p* is likely to be well adapted. On the other hand, *h*-helping has a problem too. It may induce an individual with well-adapted *p* to invest in other individuals that are related (as they have the same *h*) but happen to have poorly adapted *p*. Helping individuals with poorly adapted *p* conflicts with the notion that the survival of lineages

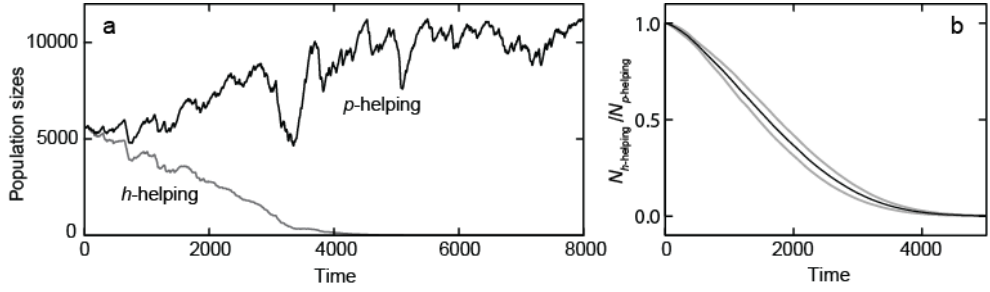

**Fig. 4 a** Two populations with different forms of fitness are made to compete (implicitly through the Malthusian factor in Eq. 4, depending on combined population size), starting with equal number of individuals in each population. The graph shows how the populations change over time, as dependent on a time-varying environment $E$. The grey curve marked '$h$-helping' implements a population that incorporates inclusive fitness (direct fitness as well as the indirect part of inclusive fitness). Helping depends on the size of $h$-subpopulations through the nonlinearity shown in Fig. 2b. The black curve marked '$p$-helping' implements a population that incorporates direct fitness as well as the phenotypic part of extensive fitness (thus without $h$-helping). Helping depends on the size of $p$-subpopulations through the same nonlinearity as used for $h$-helping. Time is given in discrete time steps of the simulation; the typical organism's lifetime equals 100 time steps. **b** Result of 100 simulations like in **a** for different realizations of $E$. The lines shows the ratio of the number of organisms in a population with $h$-helping ($N_{h\text{-helping}}$) and one with $p$-helping ($N_{p\text{-helping}}$). Black line: mean, grey lines: mean ± standard deviation.

depends on individuals that are well adapted to the environment (natural selection). As it turns out, the error made in the two cases is exactly the same (see the "Appendix", section "Misfired hereditary transfer and misfired natural selection"). Therefore, as far as the misfiring argument goes, neither $h$-helping nor $p$-helping should outcompete the other one.

The reason that $p$-helping performs better than $h$-helping can be understood from the snapshot of Fig. 2a (which is qualitatively similar to simulations with either $h$-helping or $p$-helping). The faster adaptation of $b$ makes the distribution of values $p=h+b$ in the population follow $E$ more closely than the distribution of values $h$ in the population. Both distributions contain the same individuals, and therefore the same number of individuals, but $p$ is focussed more closely on $E$ than $h$ is. The distribution of $p$ is therefore narrower and higher. Higher is the decisive property here, because subpopulations with larger numbers of individuals obtain a fitness benefit (Fig. 2b). In other words, $p$-helping outcompetes $h$-helping because subpopulations based on phenotype are typically larger than subpopulations based on heredity, and helping in large groups is more effective than helping in small groups.

Note that this result depends on how strongly $h$ and $b$ can influence $p$. One way to vary that would be to give different weights to $h$ and $b$, but here it is, equivalently, implemented in the range over which $b$ can vary (see the "Appendix", below Eq. 6). This range was chosen here to allow both $h$ and $b$ to influence the result about equally much. If $b$ could vary only negligibly, $p \approx h$, which reduces the model to one with hereditary change only. If $b$ is allowed to vary without limit, the model produces no hereditary change at all, because the faster timescale of changing $b$ compared with $h$ implies that only changing $b$ produces higher fitness than also changing $h$.

Finally, $h$-helping and $p$-helping can be combined, implemented here by weighting the $h$- and $p$-subpopulations and multiplying their respective fitness increases (see the "Appendix", section "Inclusive and extensive fitness"). Full weight on $h$ then produces $h$-helping as a special case, and full weight on $p$ produces $p$-helping. For intermediate weights, populations generally outcompete populations with only $h$-helping or $p$-helping, with highest fitness when weighting is approximately balanced between $h$ and $p$. Figure 5 shows this for competitions between a population with only $p$-helping and a population with combined helping by 50% of the members of an organism's $h$-group and by 50% of the members of its $p$-group. The latter population outcompetes the former. This is expected given the implementation, because the total fitness benefits come from two different groups ($h$ and $p$). This is beneficial with a compressive nonlinearity as in Fig. 2b, because for such a nonlinearity two approximately equal groups work better than one group twice as large. But also for models that are more realistic, it seems likely that some combination of $h$- and $p$-helping will perform better than pure

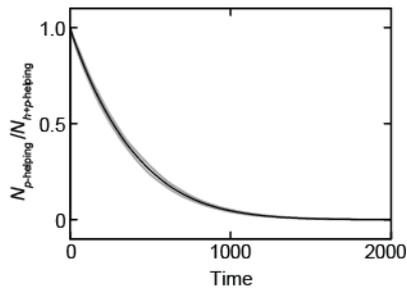

**Fig. 5** Populations with combined *p*- and *h*-helping outcompete populations with only *p*-helping. The lines show the ratio of the number of organisms in a population with *p*-helping ($N_{p\text{-helping}}$) and one with *h*-helping from 50% of the members of an organism's *h*-group and *p*-helping from 50% of the members of its *p*-group ($N_{h+p\text{-helping}}$), as a function of time (100 time steps correspond to an organism's lifetime). Black line: mean of 100 simulations with different realizations of the environment $E(t)$, grey lines: mean ± standard deviation

*p*-helping, because of the increased stability this may produce. Defence against imposters is easier for *h*-helping than for *p*-helping, that is, it is easier to detect non-*h* pretending to be *h* than it is to detect non-*p* pretending to be *p*. In practice, *h*-helping and *p*-helping might interfere in some way, affecting the results. For example, interests of members of the *h*-subpopulation may conflict with those of the *p*-subpopulation ('relatives' compete with 'friends', similarly as direct fitness may conflict with indirect fitness, as in sibling conflict, Trillmich and Wolf 2008). However, I did not attempt to model that.

The extension of *h*-helping (inclusive fitness) with *p*-helping is called here 'extensive fitness'. Extensive fitness thus incorporates inclusive fitness, but adds the fitness based on indirect phenotypic similarity to it, just as inclusive fitness incorporates direct fitness, but adds the fitness based on indirect genetic similarity to it. Note that the term 'adds' should not be taken too literally here, because the nonlinear effects of inclusive and extensive fitness (Fig. 2b) entangle the various fitness components and make them inseparable to some extent. The current approach can therefore not be formalized as separate evolution of hereditary and phenotypical components, as in gene-culture coevolution (Boyd and Richerson 1985; Richerson and Boyd 2005). Moreover, a major difference of the present theory with all previous theories – including ones that look similar, such as the general fitness as defined by Barton (2008) – is that it includes $f_{est}$, and hence agency of the individual organism. The crucial role of the individual's agency (see also below) makes this aspect of the theory quite distinct from approaches that exclusively take non-agential entities – such as genes, memes, and groups – as the carriers of evolutionary change.

## Discussion

The modelling in the previous section shows that part of fitness can be based on phenotypic similarity (including the effects of behavioural plasticity) rather than hereditary similarity. It is consistent with 'like begets like' as a basic condition for evolution by natural selection, where the 'like' can be produced by any means available. However, *p*-based fitness can only work when several conditions are fulfilled, three of which are of major importance. The first condition is that organisms should be capable of flexibly changing their own behaviour. Their behavioural repertoire must readily adapt over a considerable range to partly unpredictable variations in the environment, all in real-time during the organism's lifetime. In the variable environments assumed here, only such fast and effective adaptation can lead to *p*-groups that are significantly larger than *h*-groups. In non-variable environments, adaptation would be simpler and more stable through modified heredity rather than through behavioural change on top of fixed heredity.

The second condition for *p*-based fitness is that organisms should be capable of reliably recognizing variable behaviours in others. Because variations are partly unpredictable and occur on the fast timescale of within-lifetime behavioural changes, recognizing them is inherently difficult. Recognizing phenotypic behaviour *p* without making too many mistakes presumably requires more cognitive resources than recognizing *h*. The latter is easier, because phenotypic traits fixed by heredity are constant over an organism's lifetime. In general, kinship is indicated by proximity, the fact that organism often remain close to where they were born. But kinship is also indicated by a range of phenotypic features that are reliable and not easily faked when occurring in combination (for example, bodily colours, smells, general build, idiosyncrasies of locomotion, and so on). This implies that opportunities for impostors and freeloaders are larger for *p* than for *h*. Developing freeloading for *p*-helping can occur much faster than for *h*-helping, through behavioural plasticity. Recognizing freeloaders will thus take more cognitive resources for *p* than for *h*. Defense against freeloading requires a range of mechanisms such as those discussed elsewhere in the literature. Here it is merely assumed that this problem can indeed be solved sufficiently well.

The final condition for $p$-based fitness is that organisms should be capable of readily transferring behaviours to others and copying them from others. Moreover, the helping behaviour should extend beyond the typical reproductive setting of $h$-helping. Extensive fitness is not just reproduced heredity, but fitness in the form of any phenotypic dissemination. Only humans appear to be capable of systematically disseminating their phenotype, such as through $p$-helping (Tomasello and Carpenter 2007; Tomasello 2009), although the phenomenon may be present in rudimentary form in other species as well (de Waal 2008). Humans also fulfil the first two conditions, by having very flexible behavioural repertoires and having sufficient cognitive resources to recognize such repertoires in others. Therefore, the theory presented here is, presumably, primarily applicable to humans.

The effect of $p$-helping appears to imply a form of cultural, extra-genetic inheritance. However, the model contains no explicit inheritance beyond the transmitted heredity $h$. Explicit cultural or social inheritance, directly transmitting $b$ to $b$, is absent from the model. The appearance of cultural transmission is therefore an emergent phenomenon, produced by how the model is affected by $f_{est}$ and by how $f_{est}$ depends on both $h$ and $b$. Nevertheless, this emergent phenomenon can be enhanced by co-evolving mechanisms specifically enabling cultural transmission. Such mechanisms, and their effects on regular fitness, have been studied extensively (e.g., Boyd and Richerson 1985; Jablonka and Lamb, 2005; El Mouden et al., 2014). They could be included in future elaborations of the current model, and would have the effect of further strengthening the case of extensive fitness. The specific human capacity for anticipatory, cooperative reasoning (Kabalak et al. 2015) would have a similar effect. The evolutionary advantage of $p$-helping can be seen as a prerequisite for cooperation being rational. Rationality presumably depends critically on the presence of $f_{est}$ (van Hateren 2014b).

How extensive fitness could have arisen during hominin evolution is not addressed by the model presented here, and that question is therefore, strictly speaking, beyond the scope of this article. Nevertheless, the computational structure suggests a possible, albeit speculative, evolutionary route. Because $p$-helping can be seen as a generalization of $h$-helping (as part of inclusive fitness), a possible chain of events may have been as follows. A first step would have been the evolution of enhanced $h$-helping, as enabled by an increased accuracy of kin recognition and an increased accuracy of interpreting their behaviour. Such an enhancement may have been mutually reinforced by co-evolved changes in the social structure of hominin groups – e.g., increases in group size (Dunbar 1998), changes in child rearing (Burkart et al. 2009; Hawkes 2014), and changes in family structure (Chapais 2008). Once enhanced cognition had led to enhanced $h$-helping, this could generalize to $p$-helping based on such cognition. Such a generalization presumably co-evolved with adaptations that specifically take advantage of wide-ranging cooperation – e.g., defence against predators, further enhanced child rearing, stable and diverse acquisition of food despite environmental variability (deMenocal 2011), and adaptations enabling cumulative culture (Herrmann et al. 2007), in particular tool use and language.

Extensive fitness is a genuine fitness in the sense of quantifying a production rate, in this case not the production of hereditary 'likes', but the production of phenotypic 'likes' – similar through heredity or behaviour, or a combination. However, producing phenotypic 'likes' is, from a statistical point of view, dependent on regular direct fitness, because it can only function as a nonlinear enhancement (through $p$-helping) of the basic direct fitness. This is not really different from the indirect part of inclusive fitness, which is also dependent on direct fitness, again through a nonlinear enhancement ($h$-helping) of the basic direct fitness. Although individuals can acquire inclusive and extensive fitness independently of their own direct fitness, for a population as a whole, inclusive and extensive fitness can only work when there is sufficient direct fitness. Nevertheless, both inclusive and extensive fitness can considerably boost the fitness of the organisms in a population, with fitness understood in the general sense as the capability to produce similar organisms. The enhanced fitness subsequently increases the likelihood of evolutionary success – of the organisms and thereby of the population.

In effect, extensive fitness utilizes two forms of memory. First, the conventional memory of inclusive fitness, through heredity on evolutionary timescales and through behavioural change (phenotypic plasticity) on individual timescales. Second, there is an emergent population-based memory. When new individuals are born into the population, they already find a population that is structured to be reasonably well adapted to $E$ (Fig. 2a). Highly populated subpopulations of type $p$ will have an enhanced $f_{\text{true}}$ (because of Fig. 2b), and therefore an enhanced $f_{\text{est}}$. The *A* loop of Fig. 1b then tends to drive the behaviour of the new members into the direction of those popular and successful phenotypes: parts of phenotypic space with high $f_{\text{est}}$ are preferentially occupied. This is in fact similar to combining implicit versions of the conformist and model-based (prestige and success) biases as proposed by Richerson and Boyd (2005, pp. 120-126). The population structure in effect functions as a memory. It is still coupled to $f_{\text{true}}$, because the nonlinearity of Fig. 2b can only lead to high fitness when it amplifies a fitness that is reasonably high to start with. The current mechanism is therefore in principle evolutionary stable.

Nevertheless, the mechanism depends for its stability on how well $f_{\text{est}}$ estimates $f_{\text{true}}$. As extensively argued elsewhere (van Hateren 2014b, 2015a), the form of $f_{\text{est}}$ embodies what the organism judges to be important, implicitly or explicitly, for its $f_{\text{true}}$. It is also conjectured to be the prime factor involved in conscious agency. As this agency can subsequently modify $f_{\text{true}}$, the result is a complex feedback loop ($f_{\text{est}}$ affects $f_{\text{true}}$ and $f_{\text{true}}$ affects $f_{\text{est}}$). The combination of highly flexible goals in humans (i.e., an $f_{\text{est}}$ with a highly flexible structure of sub-goals) with such a feedback loop induces the risk that $f_{\text{est}}$ will lose track and will start to deviate significantly from $f_{\text{true}}$. However, a propensity for long-term deviation would have led to decline and extinction in the past. It is therefore likely that the human species has evolved checks and balances that are sufficiently sophisticated to prevent $f_{\text{est}}$ from going too far astray, at least on average.

As mentioned above, fitness can suffer from internal conflicts. This is true for inclusive fitness, where parent-offspring and sibling conflict (Trivers 1974; Trillmich and Wolf 2008) show that direct and indirect genetic fitness need not be aligned. A similar internal conflict can arise in extensive fitness. In that case, the consequences of phenotypic similarity may conflict with the consequences of hereditary similarity. Although the resulting behavioural pattern may then appear to be maladaptive from the point of view of inclusive fitness (which favours similar $h$), it may actually be adaptive from the more general point of view of extensive fitness (which favours similar $p$). For example, an individual's sacrifice for a phenotypic group may reduce the inclusive fitness but increase the extensive fitness of that individual. The reason is that although the sacrifice directly reduces the prospects of that individual's $h$, it may strongly increase the prospects of that individual's $p$. Then other individuals of the $p$-group are likely to help the $h$-type of the individual, indirectly and on average, which makes this strategy evolutionarily viable.

Just like inclusive fitness can be partitioned into direct effects (direct production of genetically similar individuals) and indirect effects ('producing' genetically similar individuals by helping genetically related individuals), this can be done for extensive fitness as well. Direct extensive fitness involves the direct production of phenotypically similar individuals, often by increasing the probability of being imitated. Examples that can at least partly be interpreted as direct extensive fitness are raising children, teaching, acting as a role model, and helping strangers if that increases the probability that they will become more similar to oneself. A major difference with mere direct inclusive fitness (i.e., direct fitness) is that the number of individuals affected is, potentially, much larger and not limited to a next generation. In the simple model presented here, it is obviously not possible to have many varieties of direct extensive fitness. However, it is present in the fact that participating in a particular $p$-group enhances the fitness of such $p$ (through the nonlinearity of Fig. 2b) and thereby enhances the attractiveness of $p$ for individuals who are not yet $p$. This occurs because of the dynamics of the *A* loop of Fig. 1b, which in effect attracts phenotypes towards high fitness variants. Note that the presence of an internalized version of fitness, $f_{\text{est}}$, as driving behavioural variability, is essential for this mechanism to work. Because the *A* loop represents agency, agency can be viewed as enabling the mechanism.

Indirect extensive fitness works by supporting individuals who are perceived as belonging to an in-group (i.e., the $p$-group in the model). Such individuals are judged similar to oneself. When one helps them to increase their extensive fitness – by increasing the likelihood that they produce more individuals like themselves (i.e., phenotypically similar) – one indirectly increases one's own extensive fitness. Direct reciprocation is not necessary, as long as the benefits to one's indirect extensive fitness overrule the costs. Examples that can at least partly be interpreted as indirect extensive fitness are helping kin and friends, helping culturally related strangers (or at least strangers than appear similar to oneself in significant ways), and contributing to social and cultural special interests groups. Again, the size of the group relevant for indirect extensive fitness can be much larger than the group typically relevant for the indirect parts of inclusive fitness. In the model, enhancing the extensive fitness of other members of one's *p*-group is directly implemented through the nonlinearity of Fig. 2b: by being a member of the group and interacting with its members, the extensive fitness of the group members increases. Note that again the presence of $f_{est}$ plays an important role. It is the primary mechanism through which evolutionary pressure can produce change on the short timescales of behaviour, rather than only on the long timescales of heredity.

Direct extensive fitness implies competition, because it involves attracting other individuals to become like *p*, and this attraction is done in competition with non-*p*. Indirect extensive fitness, on the other hand, implies cooperation, because it involves helping other (usually non-related) individuals of the *p*-group. For the one-dimensional model used here, the in-group and out-groups have very sharp boundaries (just different *p*). There is neither competition within *p*-groups nor collaboration between different *p*. However, realistic models would be much more complex. Humans would simultaneously belong to many different groups with varying extent, and groups would be far from homogeneous. However, the two main points that are already clear from the current model are likely to remain valid. Firstly, cooperation between non-relatives is at least partly a consequence of the presence of extensive fitness. Secondly, the balance between competition and cooperation depends on the balance between the direct and indirect parts of extensive fitness.

It should be noted that the problem of how cooperation could evolve and how it can remain stable is not so much solved by the current formulation, but rather transformed. Stabilizing mechanisms are still needed (e.g., Gächter et al. 2008; reviewed in Rand and Nowak 2013). Yet the transformation induces new, additional interpretations of well-known phenomena. For example, behaviour that increases one's reputation and status is conventionally interpreted as contributing to inclusive fitness, e.g., by increasing the probability of obtaining mates or by increasing the future probability of being helped. These are forms of direct extensive fitness derived from hereditary benefits. But in addition, such behaviour can be interpreted as enhancing direct extensive fitness phenotypically, because high status and a good reputation will increase the probability of others attempting to become similar in phenotype (independent of their heredity).

## Conclusion

As argued here, fitness for the human species needs to be extended with a component determined by groups of phenotypically similar individuals rather than being limited to genetic similarity (inclusive fitness). The result, extensive fitness, contains inclusive fitness as a special case. Extensive fitness can work because the standard external fitness is assumed to be accompanied by an internal fitness estimate, $f_{est}$. This estimate can stochastically drive behaviour into directions that, on average and in a probabilistic sense, are expected to be beneficial. In particular, it can support behaviour that increases fitness by mutual helping within phenotypically defined groups.

Extensive fitness has direct and indirect components. Direct extensive fitness is analogous to (and includes) direct genetic fitness (i.e., producing offspring). It incorporates all mechanisms by which an individual can induce others to become more like that individual. Indirect extensive fitness is analogous to (and includes) the indirect parts of inclusive fitness (i.e., helping genetically related individuals). It incorporates all mechanisms by which an individual can help other, already similar individuals to increase their extensive fitness.

The balance between direct and indirect inclusive fitness roughly corresponds to the balance between competition and cooperation. Although cooperation thus forms a basic part of human fitness, potential conflicts between different aspects of extensive fitness still require a range of special mechanisms in order to control the adverse side effects of cooperation, such as cheating and freeloading.

## Appendix

The model is an extension of the model of figure 3a and section 4.5 in van Hateren (2015a), for more details see there. A population is given by $n(h,b,t)$, with *n* the number of individuals of hereditary type *h* and behaviour *b* at time *t*. For simplicity, *n* is taken as a continuous variable and *h* and *b* are restricted to integer values. The dynamics is based on a first-order differential equation that, for a case without hereditary change, would read

$$\frac{dn}{dt} + n(1 - f_{\text{true}})/\tau_H = 0, \quad (1)$$

with $\tau_H$ the typical lifetime of an individual (that is, $1/\tau_H$ the rate of dying), and $f_{\text{true}}$ the rate of reproduction normalized over the typical lifetime. Then $f_{\text{true}}=1$ corresponds to the replacement level (balanced birth and death rates) that keeps $n$ stable, whereas $f_{\text{true}}>1$ implies exponential growth and $f_{\text{true}}<1$ decline. Hereditary change is modelled as a convolution along the $h$-dimension with a weighting function $\Lambda_H$

$$\frac{dn}{dt} + [n - \Lambda_H * (nf_{\text{true}})]/\tau_H = 0. \quad (2)$$

The convolution spreads the types of offspring of a parent of type $h$ to neighbouring types. The width of the weighting function (taken here as a normalized Gaussian with standard deviation $\lambda_H$) is therefore similar to a mutation rate (i.e., a rate of $\lambda_H/\Delta t$, with $\Delta t$ the time step of the simulation). The width is assumed to be not a constant, but a function of an estimate of $f_{\text{true}}$ made in the organism itself, i.e., $\Lambda_H(f_{\text{est}})$ and $\lambda_H(f_{\text{est}})$. Moreover, neither $f_{\text{true}}$ nor $f_{\text{est}}$ are constants, but rather processes – modelled as mere functions in this toy model – depending on the time-varying properties of organism and environment $E$, giving

$$\frac{dn(h,b,t)}{dt} + [n(h,b,t) - \Lambda_H(f_{\text{est}}(E,n,h,b)) * (n(h,b,t) f_{\text{true}}(E,n,h,b))]/\tau_H = 0. \quad (3)$$

Fitness is given by a Gaussian centred on $h+b=E$, that is, maximal when the phenotype of the organism matches the environment

$$f_{\text{true}}(E,n,h,b) = f_{\max} \exp(-(h+b-E(t))^2/2\sigma_E^2) \exp(-N(t)/K), \quad (4)$$

with $N(t) = \sum_{h,b} n(h,b,t)$ the total population size, producing a Malthusian fitness reduction (with $K$ the carrying capacity of the environment), $f_{\max}$ the maximum fitness, and $\sigma_E$ determining the width of the fitness function. The Malthusian factor in effect produces competition between the individuals of a population, as well as between (the individuals of) two different populations sharing the same environment and the same $K$. In the latter case, $N(t)$ is the total of the two populations.

For all calculations $\tau_H=100$, $f_{\max}=3$, $\sigma_E=5$, and $K=10000$ are used (units in discrete time steps $\Delta t$ and units of $h$). The value of $K$ is only nominal, because $n$ is taken as continuous and therefore the dynamics is that of an effectively infinite population size. Control calculations simulating finite population sizes by adding Poisson noise to $n$ showed that the conclusions of the article are not critically dependent on population size, unless it becomes quite small (see below). Equations (3) and (5) were evaluated through autoregressive filtering (van Hateren 2015a) performed in R (sources of the simulations available at https://sites.google.com/site/jhvanhateren/ or upon request from the author).

The conditional probability $P(b|h,t)$ of behaviour $b$ given a particular type $h$ at time $t$ is generated again by first-order dynamics as

$$\frac{dP(b|h,t)}{dt} + [P(b|h,t) - \Lambda_B(f_{\text{est}}) * P(b|h,t)]/\tau_B = 0, \quad (5)$$

where $\tau_B$ (=1 in the calculations) is the typical lifetime of a behaviour. The asterisk denotes convolution along the $b$-coordinate with a normalized Gaussian $\Lambda_B$, of which the standard deviation $\lambda_B$ is proportional to the rate of behavioural change ($\lambda_B/\Delta t$). The dynamics of Eq. (5) lies between the two following boundary cases. If $\Lambda_B$ is a $\delta$-function (no behavioural change), then the term between square brackets in Eq. (5) equals zero, implying no change in $P(b|h,t)$. On the other hand, a uniform $\Lambda_B$ implies that the convolution produces a uniform probability density as well, implying that $P(b|h,t)$ will move towards uniformity with a relaxation time $\tau_B$.

Equations (3) and (5) are coupled through

$$n(h,b,t+\Delta t) = P(b|h,t) n(h,t) = P(b|h,t) \sum_b n(h,b,t), \quad (6)$$

which continually updates the distribution of behaviour over each type $h$ in the population. For individuals, this would correspond to updating behaviour by stochastic realizations of the probability density $P(b|h,t)$. The allowed range of $b$ is taken as $-b_{\max}..b_{\max}$, with $b_{\max}=10$; values beyond the allowed range are replaced as in circular convolution. The results are qualitatively insensitive to the exact value of $b_{\max}$, with $p$-helping consistently outcompeting $h$-helping. However, the outcome of such competing becomes uncertain when $b_{\max}$ is made small in combination with small population sizes (when selection randomness, which is similar to genetic drift, is approximated by continually adding Poisson noise to the population distributions). Thus for *p*-helping to outcompete *h*-helping, both behavioural flexibility and population size need to be fairly large.

The rates of hereditary and behavioural change are not fixed, but taken to be approximately inversely proportional to a fitness estimate $f_{est}$ (Fig. 1b). Empirically, the following functions were found to produce populations that prosper under the conditions of $E(t)$ and that would typically outcompete other populations (van Hateren 2015a)

$$\lambda_H(f_{est}) = 3.6/(f_{est} + 0.1) \qquad (7)$$
$$\lambda_B(f_{est}) = 0.9/(f_{est} + 0.1). \qquad (8)$$

The $f_{est}$ in either of these expressions need not be completely identical to one another, because the factors that are taken to be most important for estimating fitness may be different on an evolutionary timescale (as relevant for an entire line of descent of organisms) from those taken to be important on a behavioural timescale (as relevant for an individual organism). However, for simplicity, they are both assumed to be identical to $f_{true}$ in the computations made here. In other words, the organisms can make perfect estimates of Eq. (4). This simplification is not crucial, though, because the mechanisms still work when estimates are not perfect (see van Hateren 2015a, sections 3 and 4.2 for discussions).

The environmental variable $E(t)$ is taken as filtered Gaussian white noise, using as filter a normalized sum of low-pass filters, with pulse response (for $t \geq 0$; for $t < 0$ $h(t)=0$) given by

$$h(t) = \sum_{i=1}^{k} \frac{a_i}{\tau_i} e^{-t/\tau_i}, \qquad (9)$$

with

$$\tau_i = r^{i-1} \tau_z \qquad (10)$$
$$a_i = \tau_i^q \Big/ \sum_{i=1}^{k} \tau_i^q. \qquad (11)$$

The noise is generated with a standard deviation of 1000 per time step, and the filter parameters are $\tau_z=4$, $r=4$, $k=6$, and $q=1$. The resulting $E(t)$ was offset to positive values for convenience, and it has approximately a power-law power spectrum (figure 4 in van Hateren 2015a). Power-law spectra are ubiquitous in nature (e.g., Bell 2010), indicating structure distributed over a wide range of timescales.

**Inclusive and extensive fitness**

For extending the model to inclusive and extensive fitness, a helping pair in a subpopulation of identical *h* or identical *p=h+b* is assumed to obtain a fitness gain by multiplication with a factor *g*. With *n* individuals in a subpopulation (either $n_h(t) = \sum_b n(h,b,t)$ or $n_p(t) = \sum_{h,b|h+b=p} n(h,b,t)$), the fitness is multiplied by $g^{\log(n)}$ (with $n \geq 1$), where the (natural) logarithm ensures that fitness rises more slowly for large subpopulations (there is a limit to how completely an individual can engage with all members of its group; for $n=1$ the factor equals one, for $n=2$, i.e., one helping pair, the factor is actually $g^{\log(2)}$ rather than *g*). Because $g^{\log(n)} = n^{\log(g)}$ (as can be seen by taking the logarithm), the fitness gain is simply a power function of *n* (as illustrated in Fig. 2b). For the calculations, *g* was set at 1.05, but the results are qualitatively insensitive to its exact value. Combining *h*-helping and *p*-helping into extensive fitness was implemented as a fitness gain $g^{\log(wn_p)+\log((1-w)n_h)}$ (with $0<w<1$ and the arguments of the logarithms $\geq 1$). The factors *w* and *1-w* can be interpreted as the probabilities of engaging with members of one's *p*- and *h*-subpopulation, respectively.

**Misfired hereditary transfer and misfired natural selection**

As argued in the main text, using either *h* or *p* for forming subpopulations with mutual helping both involves errors (misfiring) from an evolutionary point of view. When *h* is used for categorization, the error consists of helping an *h* where helping a *p* would be more appropriate from the point of view of selecting individuals that are well adapted (which depends on how well *p* matches the environment). Thus when an individual with $p_1=h+b_1$ helps an individual with $p_2=h+b_2$, the selection error made is $|p_1-p_2|=|h+b_1-(h+b_2)|=|b_1-b_2|$. The expectation of the error for the entire population is then $\langle |b_1-b_2| \rangle$, where the brackets denote a population average. On the other hand, when using $p$ for categorization, the error consists of helping to transfer an $h$ to the next generation that may be quite different from one's own $h$ and therefore might produce a rather different $p$ in the next generation. Thus when an individual with $p=h_1+b_1$ helps an individual with $p=h_2+b_2$, the hereditary error made is $|h_1-h_2|=|p-b_1-(p-b_2)|=|b_2-b_1|$. The expectation of the error for the entire population is then $\langle|b_2-b_1|\rangle$, which is identical to the error found for $h$-categorization. The conclusion is that the selection error $|p_1-p_2|$ made in the case of $h$-helping is equal to the hereditary error $|h_1-h_2|$ made in the case of $p$-helping.